\documentclass[prd,aps,showpacs, superscriptaddress,nofootinbib, 10pt]{revtex4-1}
\usepackage[utf8]{inputenc}
\usepackage{amsmath}
\usepackage{amssymb}
\usepackage{bbold}
\usepackage{graphicx}
\usepackage{subfigure}

\newcommand{\e}{\mbox{e}}

\newcommand{\sgn}{\mbox{sgn}}
\begin{document}
 
\title{Does Geometric Coupling Generates Resonances?}

\author{I. C. Jardim}  
\email{jardim@fisica.ufc.br} \affiliation{Departamento de F\'isica,
  Universidade Federal do Cear\'a, Caixa Postal 6030, Campus do Pici,
  60455-760 Fortaleza, Cear\'a, Brazil}

\author{G. Alencar}
\email{geova@fisica.ufc.br} \affiliation{Departamento de F\'isica,
 Universidade Federal do Cear\'a, Caixa Postal 6030, Campus do Pici,
 60455-760 Fortaleza, Cear\'a, Brazil}
 
 \author{R. R. Landim}  
\email{renan@fisica.ufc.br} \affiliation{Departamento de F\'isica,
  Universidade Federal do Cear\'a, Caixa Postal 6030, Campus do Pici,
  60455-760 Fortaleza, Cear\'a, Brazil}
  
 \author{R. N. Costa Filho} 
\email{rai@fisica.ufc.br} \affiliation{Departamento de F\'isica,
 Universidade Federal do Cear\'a, Caixa Postal 6030, Campus do Pici,
  60455-760 Fortaleza, Cear\'a, Brazil}

 \begin{abstract}
 Geometrical coupling in a co-dimensional one Randall-Sundrum scenario (RS) is used to study resonances of $p-$form fields. The resonances are calculated using the transfer matrix method. The model studied consider the standard RS with delta-like branes, and branes generated by kinks and domain-wall as well. The parameters are changed to control the thickness of the smooth brane. With this a very interesting pattern is found for the resonances. The geometrical coupling does not generate resonances for the reduced $p-$form in all cases considered.
 \end{abstract}

\maketitle
\section{Introduction}
Brane-world models consider the four dimensional universe as a hyper-surface in a high dimensional bulk. In this scenario Kaluza and Klein tried to unify the 
electromagnetic and gravitational fields, using a compact extra dimension \cite{Kaluza:1984ws}. Since then, the property of fields and particles in the presence of compact extra dimensions has been studied and much is known on this subject today \cite{Appelquist:1988fh}. The four dimensional world is obtained by expanding the extra dimensional dependence of the field in Fourier modes, where each one is characterized by the integer $n$. 
For example, a massless gauge field in five dimensions can generate a four dimensional one simply by considering the zero mode $n=0$. From the action viewpoint, the integral over the extra dimension is always finite since it is compact. Therefore, it trivially gets a massless gauge invariant effective action. By the same reasoning, the extra dimensional dependence on the massive modes can also be integrated in the action, giving the tower of massive fields in four dimensions said before. Gauge invariance can be used to fix the fifth component of the five dimensional gauge field to zero and therefore free fields are trivially obtained in four dimensions.

An alternative approach to compactification was proposed by Randall and Sundrum as a way to solve the hierarchy problem \cite{Randall:1999vf}.
In brane scenarios with extended extra dimension the four dimensional zero mode is not trivially obtained. Since the extra dimension is infinity, the field must to be trapped in the membrane in order to recover the four dimensional physics. The pathway is to solve an effective Schr\"odinger-like equation for the mass modes and verifying its square integrability. It is well know that in anti-de Sitter bulk, as in Randall-Sundrum model (RS), the gravity and the scalar fields are confined to the membrane \cite{Randall:1999vf}. However, due to its conformal invariance the vector field is not localized, which is a serious problem for a realistic model. Many solution has been proposed to solve this problem. For example,  some authors introduced a dilaton coupling \cite{Kehagias:2000au}, and others like in \cite{Dvali:1996xe} proposed that a strongly coupled gauge theory in five dimensions can generate a massless photon in the brane. Most of these models introduces other fields or nonlinearities to the gauge field 
\cite{Chumbes:2011zt}. 

In the pursue for a model  without introducing new degrees of freedom, a topological mass term in the bulk has been used. But they were not able to generate a massless photon in the brane \cite{Oda:2001ux}. However, it was shown by Ghoroku et al that the addition of a mass term and a boundary mass term could solve the problem \cite{Ghoroku:2001zu}. This result was generalized to include $p-$form fields \cite{Jardim:2014vba}, and more recently it was shown that the Ghoroku model can be generated by coupling the mass term with the Ricci scalar, this mechanism was called "geometrical localization" \cite{Alencar:2014moa}. Interestingly, the same mechanism was proven to work for $p-$forms and Elko spinors \cite{Alencar:2014fga, Jardim:2014cya, Jardim:2014xla}.
A similar mechanism, with a geometrical coupling in field strength, shows to be efficient to trap the zero mode of Yang-Mills field in Ref. \cite{Alencar:2015awa}. It is important to mention that a different coupling between gravity and the gauge field in RS scenarios has first been considered in  Ref. \cite{Germani:2011cv} but localization was not fully obtained. 

After the work of RS several recent results have been developed based on the idea of thick membranes and its implications for brane-world physics \cite{Bazeia:2005hu,Bazeia:2004yw,Bazeia:2003qt,Liu:2009ve,Zhao:2009ja,Liang:2009zzf,Zhao:2010mk,Zhao:2011hg, Ahmed:2013lea, Guo:2011wr, Dzhunushaliev:2009va, Movahed:2007ps, Du:2013bx,Gremm:1999pj,Bazeia:2004dh,Bazeia:2006ef,Csaki:2000fc, German:2012rv}. The advantage of these models is that the singularity generated by the brane is eliminated. In this scenario a transfer matrix method has been proposed to analyze resonances \cite{Landim:2011ki,Landim:2011ts,Alencar:2012en}. In this direction the geometrical localization mechanism has provided analytical solutions for any smooth version of the RS scenario and for any $p-$form \cite{Alencar:2014fga} (for further analytic solutions see \cite{Cvetic:2008gu,Alencar:2012en,Landim:2013dja}). When considering resonances with delta-like branes one needs to modify the transfer matrix procedure to consider a delta-like 
singularity in $z=0$ \cite{Jardim:2014vba}. 

As showed in Ref. \cite{Alencar:2014fga}, the geometric coupling traps the zero mode of all $p-$form fields in branes with
asymptotic Randall-Sundrum behavior, while for reduced $(p-1)-$form only in special cases the zero mode are localized. It is also shown that the massive modes are non-localized.
In this work we use the transfer matrix method to analyze the existence of unstable massive modes on the brane. We compute the transmission coefficient for some $p-$forms in RS scenarios, in 
brane scenario generated by a domain-wall and generated by a kink in some number of bulk dimensions. This detailed numerical study provides some hints about the structure of resonances in the geometrical localization model. 

This paper is organized as follows: In Sec. II the geometrical localization mechanism is reviewed in order to obtain the effective potential of the Schr\"odinger's equation
for the $p-$form and the reduced $(p-1)-$ form. In the third section using this potential for RS scenario the transmission coefficient for some $p-$forms and for some dimensions is calculated.
In section four an specific smooth scenario generated by thick domain-walls is used to compute the transmission coefficient again considering different $p-$forms and dimensions of the spacetime. Different values of a domain-wall
parameter, $n$, which controls that thickness of the brane is considered. In the fifth section the same procedure of previous sections for a brane generated by a kink is performed. In the last section the results are discussed.

\section{Review of Localization of $p-$form field with geometric coupling}
In this section the geometrical localization mechanism of $p-$form fields in a co-dimension one Brane-World is reviewed  based on \cite{Alencar:2014fga}. The action used is
\begin{equation}
A=\int d^{D}x\sqrt{-g}\left[-\frac{1}{2(p+1)!}Y_{M_{1}...M_{p+1}}Y^{M_{1}...M_{p+1}}-\frac{1}{2p!}\gamma_{p}R X_{M_{2}...M_{p+1}}X^{M_{2}...M_{p+1}}\right]
\end{equation}
and the equation of motion is given by
\begin{equation}\label{motionpform}
\partial_{M_{1}}[\sqrt{-g}g^{M_{1}N_{1}}...g^{M_{p+1}N_{p+1}}Y_{N_{1}...N_{p+1}}]-\gamma_{p}R \sqrt{-g}g^{M_{2}N_{2}}...g^{M_{p+1}N_{p+1}}X_{N_{2}...N_{p+1}}=0. 
 \end{equation}
Due the geometric coupling term the divergence of the equation of motion is not trivial and provides the condition
\begin{equation}\label{divpform}
 R \e^{(D-2p)A}\partial^{\nu_{2}}X_{\nu_{2} N_{3}...N_{p+1}} + \partial\left[R \e^{(D-2p)A}X_{5N_{3}...N_{p+1}}\right]=0,
\end{equation}
where $\partial$ means a $z$ derivative and from now on all $(D-1)-$dimensional indices will be contracted with
$\eta^{\mu\nu}$. From eq. (\ref{motionpform}) we can write the following component equations
\begin{eqnarray}
&&\e^{(D-2(p+1))A}\partial_{\mu_{1}}[Y^{\mu_{1}\mu_{2}...\mu_{p+1}}]+\partial(\e^{(D-2(p+1))A}Y^{5\mu_{2}...\mu_{p+1}})-\gamma_{p}R \e^{(D-2p)A}X^{\mu_{2}...\mu_{p+1}}=0; \label{pformnu}
\\&&\partial_{\mu_{1}}Y^{\mu_{1}\mu_{2}...\mu_{p}5} -\gamma_{p}R \e^{2A}X^{\mu_{2}...\mu_{p}5}=0, \label{pform5}
\end{eqnarray}
and from the divergence equation (\ref{divpform}) we get
\begin{eqnarray}
 &&\partial_{\mu_{1}}X^{\mu_{1}...\mu_{p-1}5} \equiv \partial_{\mu_{1}}X^{\mu_{1}...\mu_{p-1}}=0,
\\&&\partial(R \e^{(D-2p)A}X^{\mu_{1}...\mu_{p-1}})+R \e^{(D-2p)A}\partial_{\mu_{p}}X^{\mu_{1}...\mu_{p}}=0. \label{tracelesspform}
\end{eqnarray}

The gauge symmetry has been broken in the $D-$dimensional action because of the coupling term. To avoid this difficulty we must divide the $p-$form in transversal and longitudinal parts as
\begin{equation}\label{defXTXL}
X_{T}^{\mu_{1}...\mu_{p}}\equiv X^{\mu_{1}...\mu_{p}}+\frac{(-1)^{p}}{\Box}\partial^{[\mu_{1}}\partial_{\nu_{1}}X^{\mu_{2}...\mu_{p}]\nu_{1}};\;\;\;\;X_{L}^{\mu_{1}...\mu_{p}}\equiv \frac{(-1)^{p-1}}{\Box}\partial^{[\mu_{1}}\partial_{\nu_{1}}X^{\mu_{2}...\mu_{p}]\nu_{1}}. 
\end{equation}
In order to obtain the equation of motion for booth parts of $p$ and $(p-1)-$forms Eqs. (\ref{pformnu}) and (\ref{pform5}) must be decoupled. As showed in Refs. \cite{Alencar:2014fga} the decoupling can be obtained using the definitions
(\ref{defXTXL}) and the divergence equation (\ref{tracelesspform}) leading to
\begin{eqnarray}
&& \e^{(D-2(p+1))A}\square X_{T}^{\mu_{1}...\mu_{p}}+\partial(\e^{(D-2(p+1))A}\partial X_{T}^{\mu_{1}...\mu_{p}})-\gamma_{p}R \e^{(D-2p)A}X_{T}^{\mu_{1}...\mu_{p}}=0, \label{XTpfull}
\\&&\Box X^{\mu_{1}...\mu_{p-1}}+\partial(R\e^{-(D-2p)A}\partial(R\e^{(D-2p)A}X^{\mu_{1}...\mu_{p-1}}))-\gamma_{p}R \e^{2A}X^{\mu_{1}...\mu_{p-1}}=0.\label{phipfull}
\end{eqnarray}
To determine the longitudinal part of $p-$form we can use the divergence equation
\begin{equation}
\partial(R \e^{(D-2p)A}X^{\mu_{1}...\mu_{p-1}})+R \e^{(D-2p)A}\partial_{\mu_{p}}X_{L}^{\mu_{1}...\mu_{p}}=0.
\end{equation}

Starting with the transversal part of the $p-$form field, we impose the separation of variables in the form $ X_{T}^{\mu_{1}...\mu_{p}}(z,x) = f(z)\tilde{X}_{T}^{\mu_{1}...\mu_{p}}(x)$ in (\ref{XTpfull}) to obtain the set of equations
\begin{eqnarray}
&&  \square \tilde{X}_{T}^{\mu_{1}...\mu_{p}}-m_{X}^{2}\tilde{X}_{T}^{\mu_{1}...\mu_{p}}=0,
\\&& (\e^{(D-2(p+1))A}f'(z))'-\gamma_{p}R \e^{(D-2p)A}f(z) =-m_{X}^{2}\e^{(D-2(p+1))A}f(z), \label{eqfp}
\end{eqnarray}
where the prime means a derivative with respect to $z$. To write Eq. (\ref{eqfp}) in a Schr\"odinger form we must make $f(z) = \e^{-(D-2(p+1))A/2}\psi$. The potential obtained after this transformation is
\begin{equation}\label{potp}
U(z)=\left[\frac{\alpha_{p}^{2}}{4} -(D-1)(D-2)\gamma_{p}\right]A'(z)^{2} + \left[\frac{\alpha_{p}}{2} - 2(D-1)\gamma_{p}\right] A''(z),
\end{equation}
where $\alpha_{p} = D-2(p+1)$ and we have used that $R = -(D-1)\left[2A'' +(D-2)A'^{2}\right]\e^{-2A}$. Imposing a solution for zero mode in the form $\psi \propto \e^{bA}$ we obtain that
\begin{equation}\label{gammap}
\gamma_{p} = -\frac{(D-2)-2\alpha_{p}}{4(D-1)},
\end{equation}
and $b = p$. Fixing the coupling constant $\gamma_{p}$ by (\ref{gammap}) the zero mode of transversal part of $p-$form field is localized in a brane with asymptotic Randall-Sundrum behavior. \footnote{ The potential (\ref{potp}) and the fixation (\ref{gammap}) differ from those obtained in ref. \cite{Alencar:2014fga} by a minus signal in $\gamma_{p}$. This do not change the result for the zero mode but it is important for the massive modes.}  
To compute the massive modes we must to specify the brane scenario.

For the $(p-1)-$form we will separate the variables imposing that $X_{\mu_{2}...\mu_{p}}(z,x) = u(z)\tilde{X}_{\mu_{2}...\mu_{p}}(x)$ in (\ref{phipfull}) to obtain the set of equations
\begin{eqnarray}
&& \Box \tilde{X}^{\mu_{2}...\mu_{p}}-m_{p-1}^{2}\tilde{X}^{\mu_{2}...\mu_{p}}=0, 
\\&& \left(R\e^{-(D-2p)A}(R\e^{(D-2p)A}u(z))'\right)'-\gamma_{p}R\e^{2A}u(z)=-  m_{p-1}^{2}u(z)\label{equp},
\end{eqnarray}
and as we have made for $p-$form, we will make $u(z) = (R\e^{(D-2p)A})^{-1/2}\psi$  to transform the eq. (\ref{equp}) in a Schr\"odinger-like equation with the potential 
\footnote{ Here we have corrected a typo in the potential obtained in ref. \cite{Alencar:2014fga}.}
\begin{equation}\label{potp-1}
U(z) =\frac{1}{4}\left[(\alpha_{p}+2)A'+(\ln R)'\right]^{2} -\frac{1}{2}\left[(\alpha_{p}+2)A'' +(\ln R)''\right] +\gamma_{p}R\e^{2A}.
\end{equation}

The above potential  is not simple to be analyzed in its complete form. Therefore, the study of localization must be made by the asymptotic behavior, which is
\begin{equation}
U(z) =\left[\frac{\alpha_{p-1}^{2}}{4} + (D-1)(D-2)\gamma_{p}\right]A'(z)^{2} + \left[-\frac{\alpha_{p-1}}{2} + 2(D-1)\gamma_{p}\right] A''(z).
\end{equation}
Since we have fixed the coupling constant $\gamma_{p}$ to localize the zero mode of the $p-$form field transversal part the same can not be done for the $(p-1)-$form, with the exception for the case
when $\alpha_{p-1}^{2} + 4(D-1)(D-2)\gamma_{p} = \left(\alpha_{p-1} -4(D-1)\gamma_{p}\right)^{2} $, i.e, when $\alpha_{p} = -1$; in five dimensions this condition is satisfied only by Kalb-Ramond field. 

Finally, for the longitudinal part of $p-$form field, making the separation of variables in the form $X^{\mu_{1}...\mu_{p}}_{L}(z,x) = F(z)\tilde{X}^{\mu_{1}...\mu_{p}}_{L}(x)$ we obtain, from (\ref{tracelesspform}),
\begin{eqnarray}
&& C_{0}\left(R \e^{(D-2p)A}u(z)\right)' = R \e^{(D-2p)A}F(z),\label{eqF}
\\&&\tilde{X}^{\mu_{1}...\mu_{p-1}} +C_{0}\partial_{\mu_{p}}\tilde{X}_{L}^{\mu_{1}...\mu_{p}}=0,
\end{eqnarray}
where $C_{0}$ is a constant. To solve the massive modes we must specify the brane scenario explicitly. In the following sections we will study this for some Brane-World scenarios.  

\section{The delta-like Brane Case}
\begin{figure}[h!]
 \centering
 \includegraphics[scale=0.25]{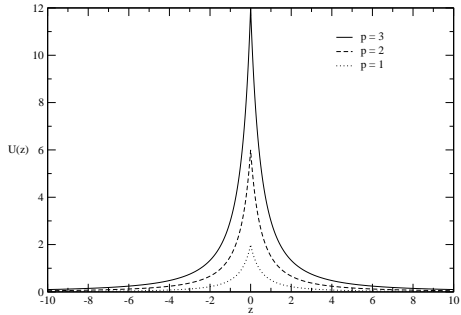}
 \caption{Plot of the Schr\"odinger potential for a $p-$form field with $p=1,2,3$ in Randall-Sundrum scenario with $k =1$.}
 \label{fig:potpRS}
\end{figure}

The first brane scenario discussed here is the one with a delta-like brane. Despite the singularity, this scenario has an historical importance and serves as an important paradigm in physics of extra dimensions and field localization. The warp factor for this scenario in a conformal form is given by
\begin{equation}
A(z) = -\ln\left[k|z| +1\right].
\end{equation}
The Schr\"odinger equation potential, Eq. (\ref{potp}),  for the transversal part of the $p-$form field is given by 
\begin{equation}\label{potpRS}
U(z)=\frac{p(p+1)k^{2}}{(k|z|+1)^{2}} -2pk\delta(z),
\end{equation}
and is illustrated in Fig. \ref{fig:potpRS} for some $p-$forms. An interesting result is that the potential of the transversal part of $p-$form does not depend on dimensionality of space-time. It depends only on the degree of the form.
As imposed in previous section the solution of zero mode is
\begin{equation}
 \psi \propto \left[k|z| +1\right]^{-p}.
\end{equation}

\begin{figure}[h!]
 \centering
 \includegraphics[scale=0.25]{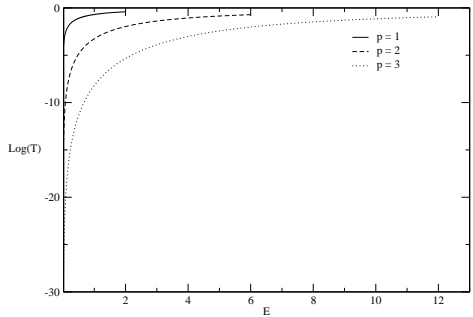}
 \caption{Transmission coefficient for $p-$form field in Randall-Sundrum scenario with $k=1$ as function of energy, $E = m_{X}^{2}$.}
 \label{fig:T-R-RS}
\end{figure}

For the massive case Eq. (\ref{eqfp}) provides the solution
\begin{equation}\label{psimaspform}
 \psi(z) =(k|z|+1)^{1/2}[C_{1}J_{\nu}(m_{X}|z|+ m_{X}/k)+C_{2}Y_{\nu}(m_{X}|z|+ m_{X}/k)],
\end{equation}
where $C_{1}$ and $C_{2}$ are constants and $\nu = (2p+1)/2$ . Now the boundary conditions impose that
\begin{equation}
C_{1} = C_{2}\frac{m_{X} Y_{\nu-1}(m_{X}/k) -2\nu k Y_{\nu}(m_{X}/k) - 
     m_{X} Y_{\nu+1}( m_{X}/k)}{ m_{X} J_{\nu-1}( m_{X}/k) +2\nu k J_{\nu}( m_{X}/k) -
   m_{X}  J_{ \nu +1}( m_{x}/k)}.
\end{equation}
Then the massive modes are non-localized. To obtain more information
about massive modes we can evaluate the transmission coefficient. For this the solution (\ref{psimaspform}) is written in the form
\begin{equation}
 \psi(z) = \left\lbrace\begin{matrix}E_{\nu}(-z)+\sigma F_{\nu}(-z) &,\;\mbox{for}\; z<0 \\ 
\gamma F_{\nu}(z)&,\;\mbox{for}\; z\geq0\end{matrix}\right.,
\end{equation}
with
\begin{eqnarray}
&&E_{\nu}(z) = \sqrt{\frac{\pi}{2}}(m_{X}z+ m_{X}/k)^{1/2}H^{(2)}_{\nu}(m_{X}z+\ m_{X}/k)
\\&& F_{\nu}(z) = \sqrt{\frac{\pi}{2}}(m_{X}z+m_{X}/k)^{1/2}H^{(1)}_{\nu}(m_{X}z+ m_{X}/k),
\end{eqnarray}
where $H^{(1)}_{\nu}$ and $H^{(2)}_{\nu}$ are the Hankel functions of first and second kind respectively. The boundary conditions at $z = 0$ imposes 
\begin{equation}
 \gamma =  \frac{W(E_{\nu},F_{\nu})(0)}{ 2F_{\nu}(0)F_{\nu}'(0) + 2pk F_{\nu}^{2}(0)},
\end{equation}
where $ W(E_{\nu},F_{\nu})(0) =  E_{\nu}(0)F_{\nu}'(0)-E_{\nu}'(0)F_{\nu}(0)$ is the Wronskian  at $z=0$. Since it is constant in Schr\"odinger equation the transmission coefficient can be written as
\begin{equation}
T = |\gamma|^{2} = \frac{m_{X}^{2}}{|F_{\nu}(0)F_{\nu}'(0) + pk  F_{\nu}^{2}(0)|^{2}}. 
\end{equation}
The transmission coefficient is plotted in Fig. \ref{fig:T-R-RS} as function of energy for some $p-$forms. The figure does not show peaks, indicating that there is no unstable massive modes.  

\begin{figure}[!h]
\centering
\subfigure[]{
\includegraphics[scale=0.25]{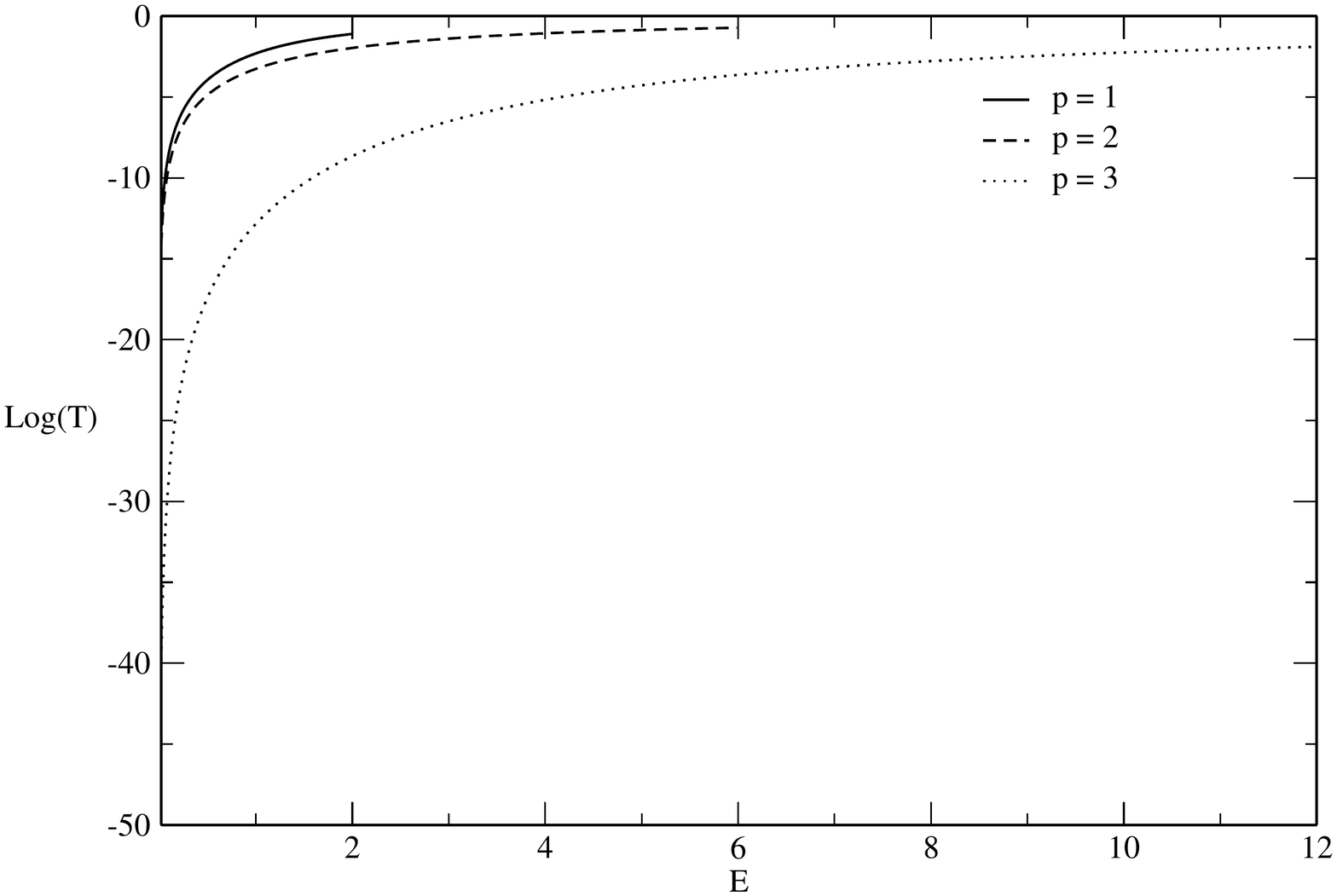}
\label{fig:TR1RS}
}
\subfigure[]{
\includegraphics[scale=0.25]{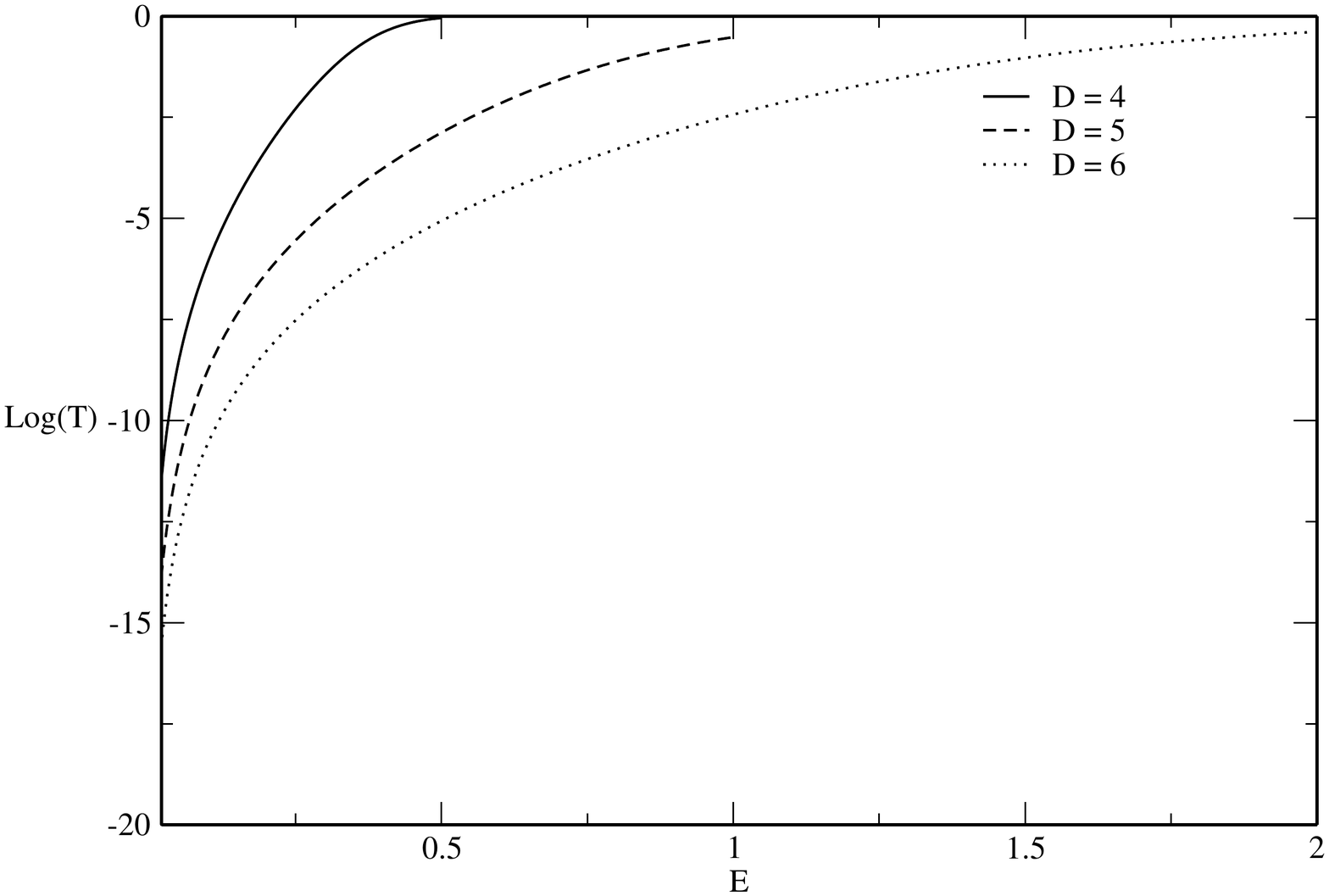}
\label{fig:TR1RSD}
}
\caption{Transmission coefficient for $(p-1)-$form in Randall-Sundrum scenario with $k=1$ as function of energy, $E = m_{p-1}^{2}$. (a) For some $p-$forms in $D=5$, (b) For $1-$form for some $D$. }
\end{figure}

For the $(p-1)-$form in Randall-Sundrum scenario the potential of Schr\"odinger equation, (\ref{potp-1}), can be written as
\begin{equation}\label{potp-1RS}
U(z)= \frac{p(p+1)k^{2}}{(k|z|+1)^{2}} -2[p -(\alpha_{p}+1)]k\delta(z).
\end{equation}
Since $R$ in RS scenario is a constant, it does not contribute to the potential. Unlike the $p-$form case, the potential of $(p-1)-$form depends on the dimensionality of the space-time. 
As a finite part of the potential is the same as in (\ref{potpRS}), the regular part of the potential and the solution are the same of the $p-$form field
\begin{equation}\label{solp-1RS}
 \psi(z) = f_{1}\left(k|z| +1\right)^{-p} +f_{2}\left(k|z| +1\right)^{p+1}.
\end{equation}
Now the boundary condition at $z=0$ imposes the connection between the constants
\begin{equation}
f_{1}\left(\alpha_{p} +1\right) + (D-2)f_{2} = 0.
\end{equation}
As discussed in the previous section, only when $\alpha_{p} = -1$ it is possible to obtain a convergent solution, i.e., a vanishing $f_{2}$. 
The potential for $(p-1)-$form is the same of $p-$form with a modified boundary condition. The behavior of massive  modes are the same, i.e, non-localized. The behavior of transmission coefficient is shown in Fig. \ref{fig:TR1RS} for some forms in $D =5$.  In Fig. \ref{fig:TR1RSD} we consider the $1-$form case in some bulk dimensions. The figure shows the same behavior as the $p-$form.

The longitudinal part of $p-$form can be found replacing the solution (\ref{solp-1RS}) in  (\ref{eqF}). This procedure provides
\begin{equation}
 F(z) = F_{1}\;\sgn(z)\left[15\left(k|z| +1\right)^{-3/2+\alpha_{p}} +(2\alpha_{p}-3)\left(\alpha_{p} +1\right)\left(k|z| +1\right)^{5/2}\right],
\end{equation}
where $F_{1}$ is a constant. As in the $(p-1)-$form case only when $\alpha_{p} = -1$ the above expression provides a localized solution.

\section{The Smooth domain-wall Brane Case}
\begin{figure}[!h]
\centering
\subfigure[]{
\includegraphics[scale=0.25]{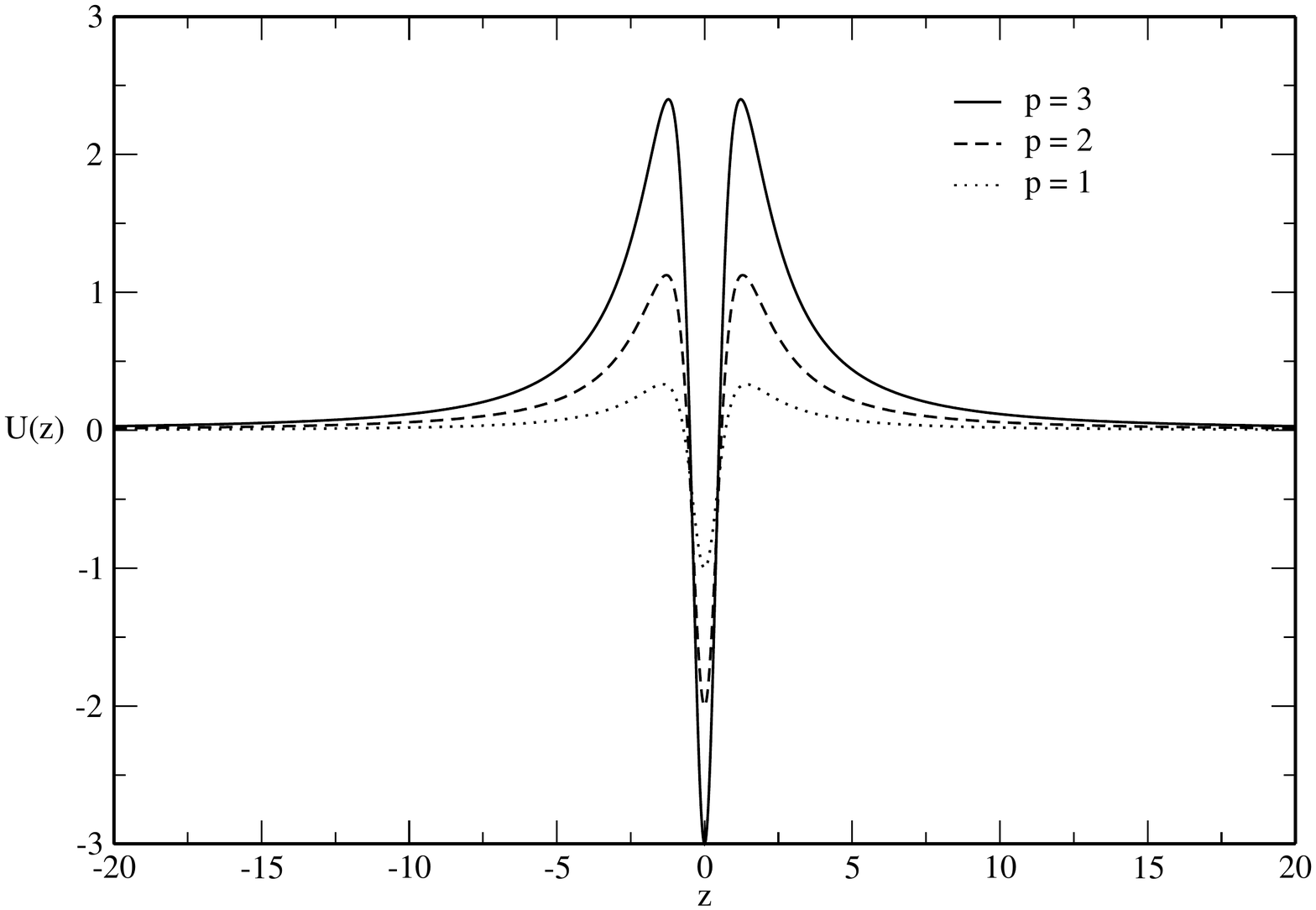}
\label{fig:ptosma}
}
\subfigure[]{
\includegraphics[scale=0.25]{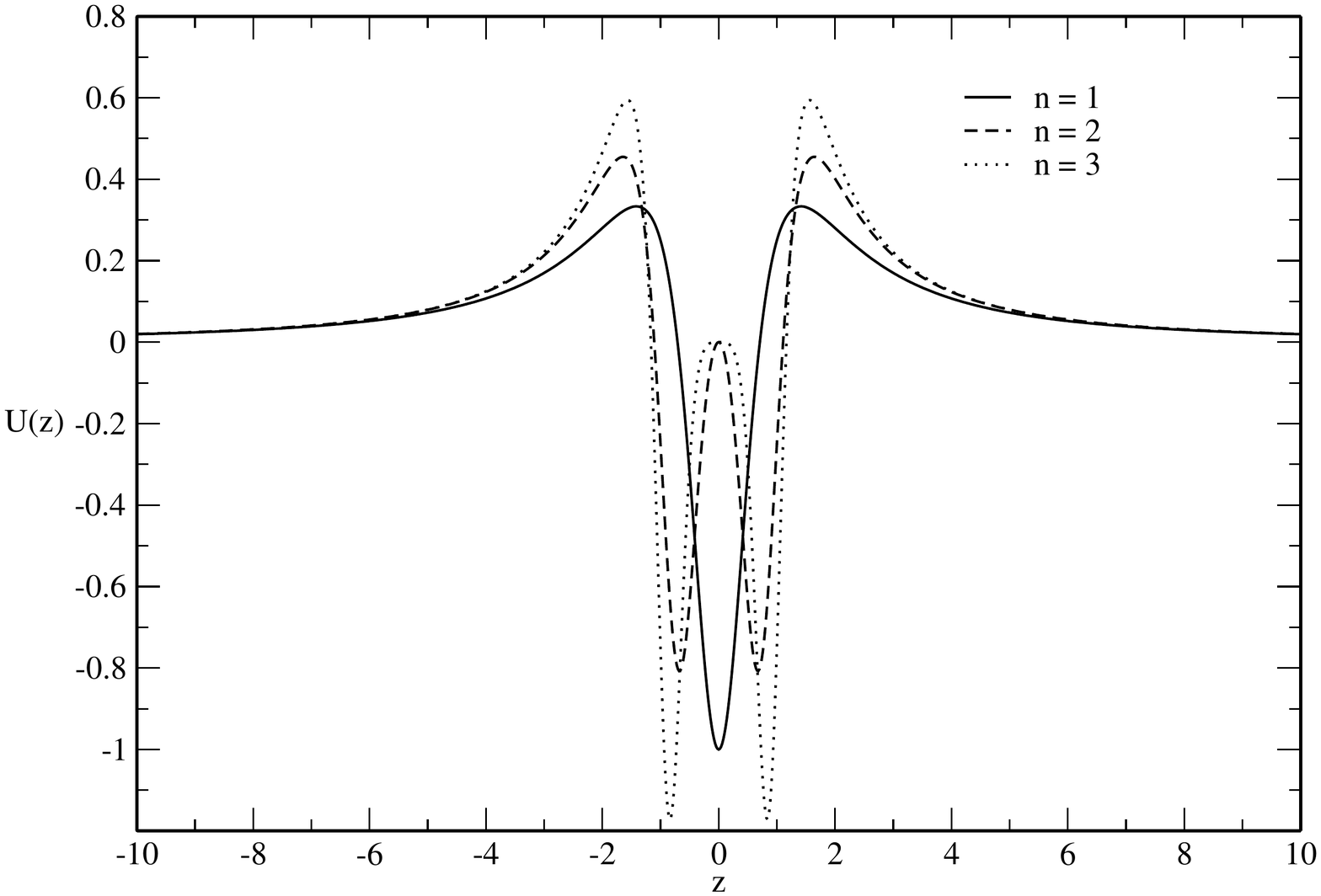}
\label{fig:ptosmn}
}
\caption{Behavior of Schr\"odinger potential in a smooth scenario generated by domain walls. (a) For some $p-$forms with $n =1$, (b) For $1-$form and for some values of parameter $n$. }
\end{figure}

In this section we investigate the localization of a $p-$form field in a smooth warp factor scenario. Since the metric can be written in a conformal form we can use all results obtained in the previous section which did not use the explicit form of the warp factor. The smooth warp factor produced by a domain-wall is used \cite{Du:2013bx, Melfo:2002wd},
\begin{equation}
  A(z) = -\frac{1}{2n}\ln\left[\left(kz\right)^{2n}+1\right],
\end{equation}
which recovers the Randall-Sundrum metric for large $z$ and $n \in N^{*}$. Using this metric in Eq. (\ref{potp}) we obtain the Schr\"odinger's potential for the transversal part of the $p-$form
\begin{equation}\label{potpsm}
U(z)=p\left(p +2n\right)\frac{k^{2}(kz)^{4n-2}}{\left[(kz)^{2n} +1\right]^{2}} -p\frac{(2n-1)k^{2}(kz)^{2n-2}}{(kz)^{2n}+1}.
\end{equation}
That is shown in Fig. \ref{fig:ptosmn} for some values of $n$ with $p =1$, and in Fig. \ref{fig:ptosma} for some values of $p$ with $n =1$.
As in the delta-like case, the potential of $p-$form does not depend on the dimensionality of the spacetime. 

\begin{figure}[!htb]
\centering
\subfigure[]{
\includegraphics[scale=0.25]{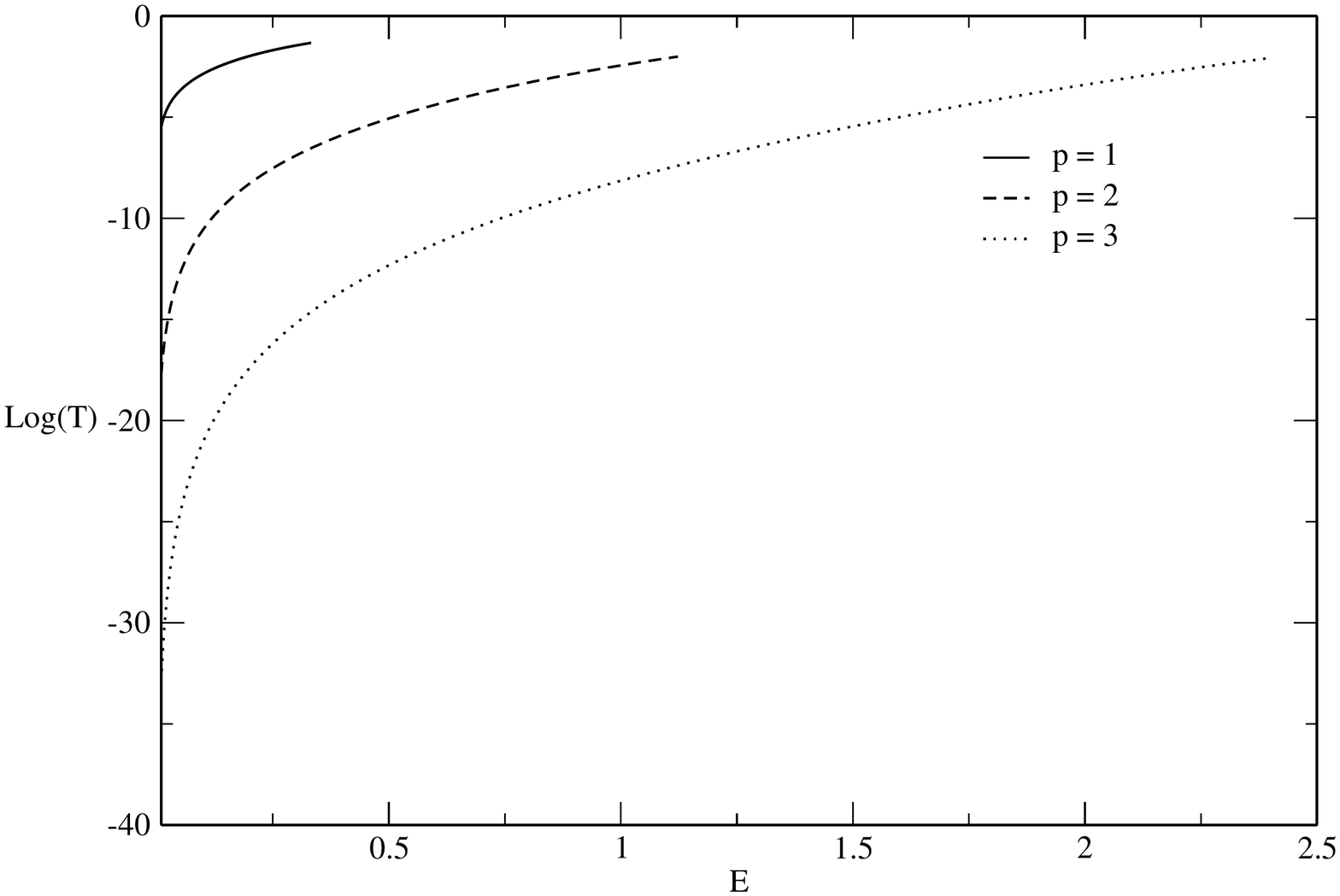}
\label{fig:T-R-smootha}
}
\subfigure[]{
\includegraphics[scale=0.25]{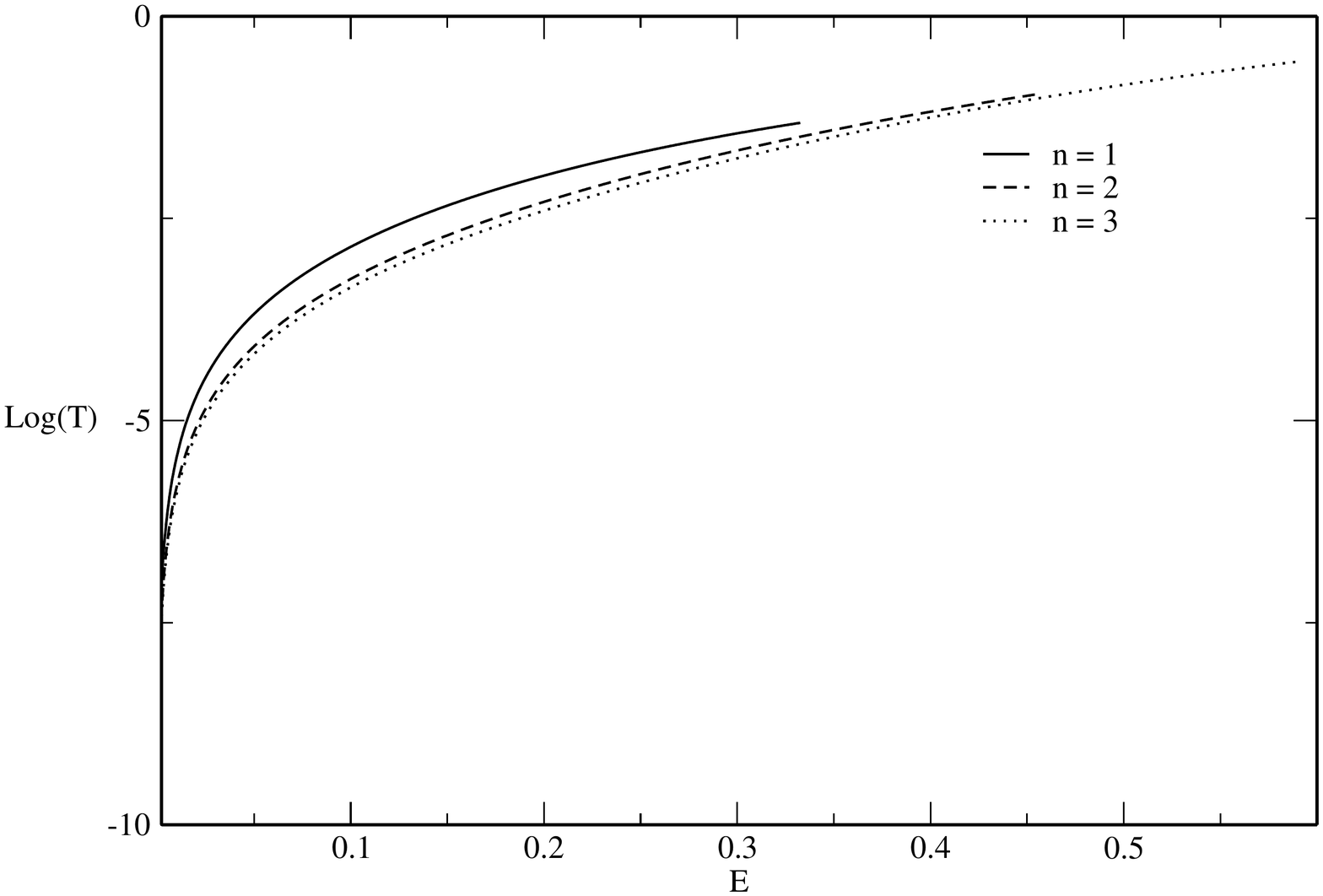}
\label{fig:T-R-smoothn}
}
\caption{Transmission coefficient in smooth scenario generated by domain walls as function of energy, $E= m_{X}^{2}$. (a) For some $p-$forms with $n =1$, (b) For $1-$form and for some values of parameter $n$. }
\end{figure}

For the massless mode of the transversal $p-$form, the Schr\"odinger's equation with the above potential provides the following convergent solution
 \begin{equation}\label{solfpsmo}
 \psi \propto \left[\left(kz\right)^{2n}+\beta\right]^{-p/2n},
 \end{equation}
as computed in Sec. II. 
The solution of massive modes of  transversal $p-$form can not be found analytically. To obtain information about these states we use the transfer matrix method to evaluate the transmission coefficient. The behavior is illustrated in Fig. \ref{fig:T-R-smoothn} for $p = 1$ and some values of parameter $n$. In Fig. \ref{fig:T-R-smootha} the transmission coefficient is shown for different $p-$forms with $n=1$. Both figures do not exhibit  peaks, indicating the absence of unstable modes.  

Considering now the massless mode of the reduced $(p-1)-$form, Eq. (\ref{potp-1}) gives a complicated potential. One must be careful since it involves $\ln R$ and a vanishing $R$ generates a divergent contribution to the potential. To deal with this one  must consider cases with a regular Ricci scalar. Its explicit form is given by
\begin{equation}\label{Rsm}
R = (D-1)z^{2(n-1)}\frac{2(1-2n) +Dz^{2n}}{(1+z^{2n})^{2-1/n}},
\end{equation}
the Ricci scalar vanishes at $z = \pm(2(2n-1)/D)^{1/2n}$, making the potential (\ref{potp-1}) divergent at these points. Because of this, the transmission coefficient can not be calculated by transfer matrix method.

The longitudinal part of $p-$form is determined by Eq. (\ref{tracelesspform}). As no analytic solution for massless mode of $(p-1)-$form is discussed here, it is not possible to find an analytic solution for massless mode of the longitudinal part of $p-$form.

\section{The Smooth Kink Brane Case} 
\begin{figure}[!h]
\centering
\subfigure[]{
\includegraphics[scale=0.25]{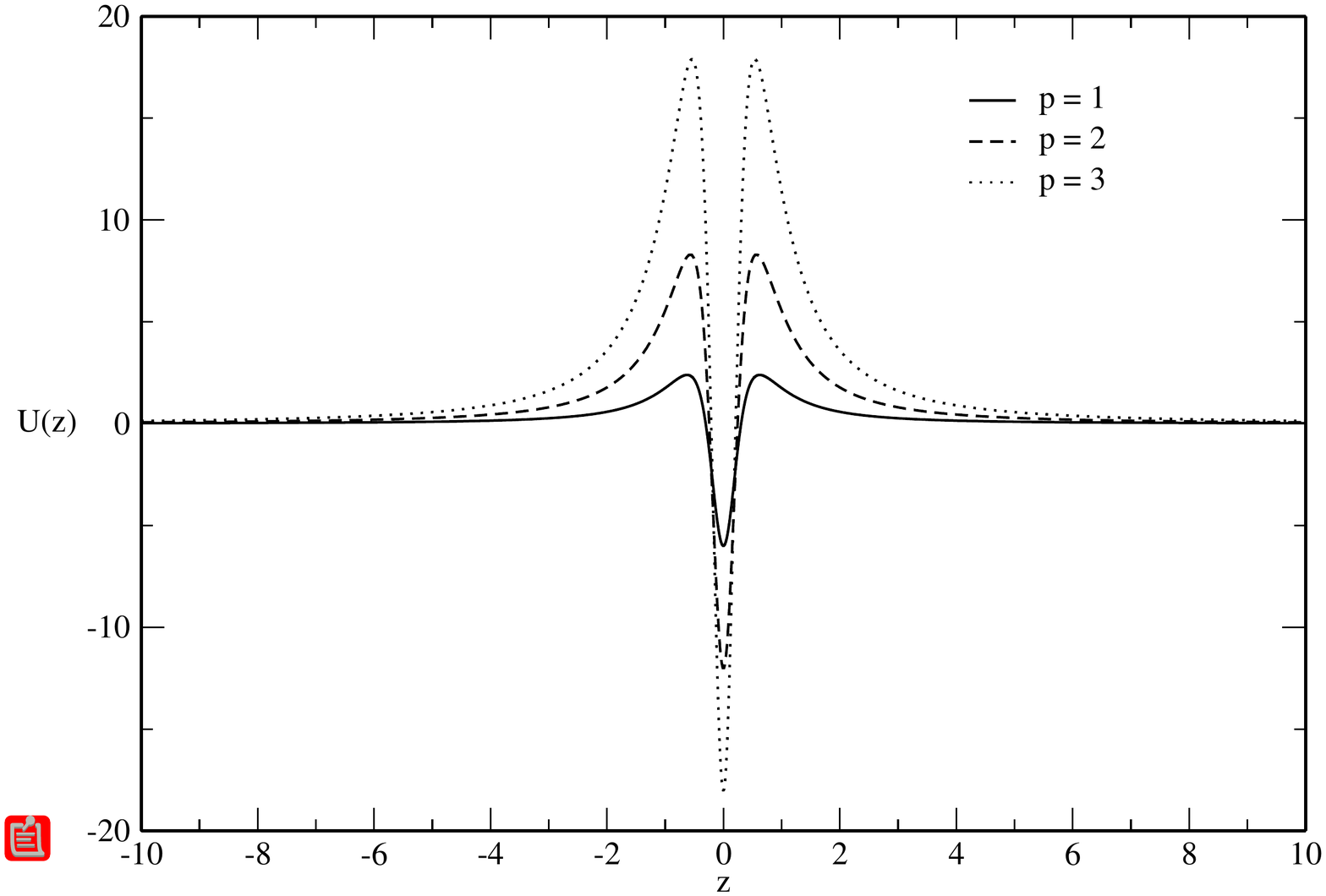}
\label{fig:potpkink}
}
\subfigure[]{
\includegraphics[scale=0.25]{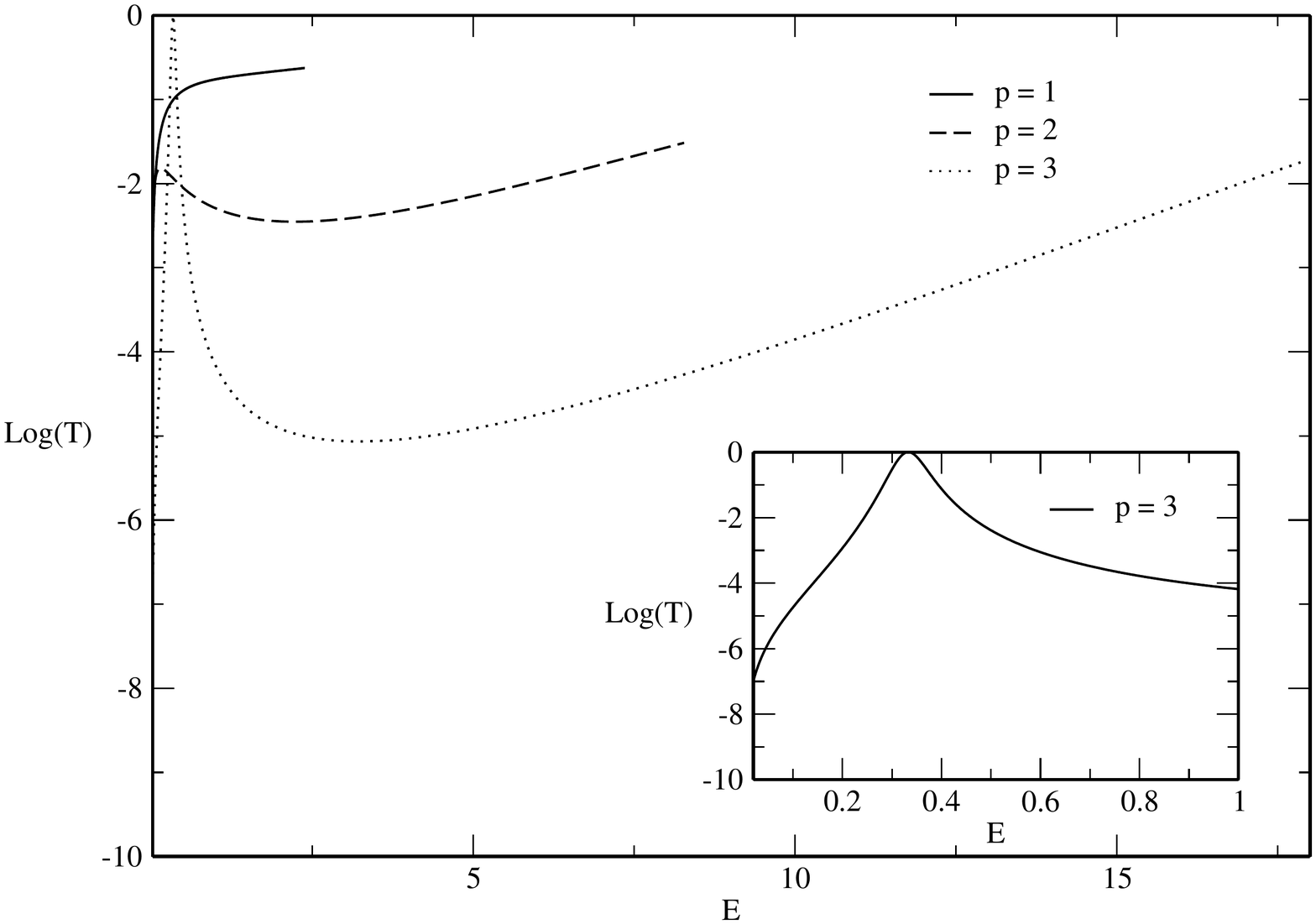}
\label{fig:TR-kink}
}
\caption{(a) Behavior of Schr\"odinger potential in kink scenario for some $p-$forms. (b) Transmission coefficient in kink scenario for some $p-$form as function of energy, $E= m_{X}^{2}$. As showed in detail, the peak for $3-$form is not a resonance.}
\end{figure}

In this section we must consider branes generated by a kink.  The warp factor is given by \cite{Landim:2011ki}
\begin{equation}\label{warpkink}
 A(y) = -4\ln\cosh y -\tanh^{2}y,
\end{equation}
where the variable $y$ is related to the conformal coordinate $z$ by $dz = \e^ {-A(y)}dy$. The behavior of the potential of the transversal part of the $p-$form field, Eq. (\ref{potp}), with this warp factor is illustrated in Fig. \ref{fig:potpkink}. For the massive modes, like in previous sections, the transfer matrix method is used to compute the transmission coefficient. The results are plotted in Fig. \ref{fig:TR-kink}, and unlike the previous cases it exhibits the most probable state when $p = 3$. A careful analyzes indicates that the peak for the $3-$form is not a resonance, as shown in detail in Fig. \ref{fig:TR-kink}. 

For $(p-1)-$form field the Schr\"odinger potential of Eq. (\ref{potp-1}) with warp factor of Eq. (\ref{warpkink}) diverges at finite points due the Ricci scalar vanishes at these same points. Because of this, the transmission coefficient can not be calculated by transfer matrix method. 
\section{conclusion}
Here we have calculated the transmission coefficient for massive modes of $p-$form fields with geometrical coupling in some brane scenarios. The important result found is that in all scenarios covered in this paper no resonances appears for the massive modes of $p-$form fields. To reach this conclusion we study a thin and two thick brane scenarios: one generated by domain walls and other generated by a kink. Since the potential to this cases do not depend on the dimension $D$, the number of parameters to be considered is fewer than for the reduced $(p-1)-$form. The case with thin or delta-like brane has also been considered previously in \cite{Jardim:2014vba} and no resonances has been found. However, there the Ghoroku model was used \cite{Ghoroku:2001zu} which contains two free parameters and is very similar to the geometrical localization mechanism if the RS warp factor is considered. Here we have consider a different range in the parameter. Fig. \ref{fig:T-R-RS} showed the cases $p=1,2,3$ and no 
resonances 
were found. The indication that there is no resonances are the plots of the potential \ref{fig:potpRS}, which have only one barrier. Next, the domain wall case were considered. For this case, the effective potential has a volcano profile fig. \ref{fig:ptosmn} and at first sight resonances should be expected. However, as showed in Figs. \ref{fig:T-R-smootha} and \ref{fig:T-R-smoothn}, for several possible parameters, no resonances appears. Finally we consider the kink case. The effective potential was plotted in Fig. \ref{fig:potpkink} and again a volcano-like potential is obtained. At this time a possible resonance appears for the $3-$form field. However, when the  numerical calculation is improved we see in Fig. \ref{fig:TR-kink} that there is no resonance at that mass. For all other cases no resonances was found. Therefore, this results indicates that the geometric coupling does no produce resonances for $p-$form fields. Since all the results presented here are from numerical calculations, more cases must 
be studied in order to reinforce that conclusion.

\section*{Acknowledgments}

We acknowledge the financial support provided by Funda\c c\~ao Cearense de Apoio ao Desenvolvimento Cient\'\i fico e Tecnol\'ogico (FUNCAP), the Conselho Nacional de 
Desenvolvimento Cient\'\i fico e Tecnol\'ogico (CNPq) and FUNCAP/CNPq/PRONEX.


\begin{thebibliography}{99}

\bibitem{Kaluza:1984ws} 
  T.~Kaluza,
  In *O'Raifeartaigh, L.: The dawning of gauge theory* 53-58
  
\bibitem{Appelquist:1988fh} 
  T.~Appelquist, A.~Chodos and P.~G.~O.~Freund,
  IN *APPELQUIST, T. (ED.) ET AL.: MODERN KALUZA-KLEIN THEORIES* 1-47

    
\bibitem{Randall:1999vf} 
  L.~Randall and R.~Sundrum,
  Phys.\ Rev.\ Lett.\  {\bf 83}, 4690 (1999)
  [hep-th/9906064].

  

\bibitem{Kehagias:2000au} 
  A.~Kehagias and K.~Tamvakis,
  Phys.\ Lett.\ B {\bf 504}, 38 (2001)
  [hep-th/0010112].

\bibitem{Dvali:1996xe} 
  G.~R.~Dvali and M.~A.~Shifman,
  Phys.\ Lett.\ B {\bf 396}, 64 (1997)
  [Erratum-ibid.\ B {\bf 407}, 452 (1997)]
  [hep-th/9612128].



\bibitem{Chumbes:2011zt} 
  A.~E.~R.~Chumbes, J.~M.~Hoff da Silva and M.~B.~Hott,
  Phys.\ Rev.\ D {\bf 85}, 085003 (2012)
  [arXiv:1108.3821 [hep-th]].
  
\bibitem{Oda:2001ux} 
  I.~Oda,
  hep-th/0103052.
  
\bibitem{Ghoroku:2001zu} 
  K.~Ghoroku and A.~Nakamura,
  Phys.\ Rev.\ D {\bf 65}, 084017 (2002)
  [hep-th/0106145].
  
\bibitem{Jardim:2014vba} 
  I.~C.~Jardim, G.~Alencar, R.~R.~Landim and R.~N.~Costa Filho,
  arXiv:1410.6756 [hep-th].
  
  
\bibitem{Alencar:2014moa} 
  G.~Alencar, R.~R.~Landim, M.~O.~Tahim and R.~N.~Costa Filho,
  Phys.\ Lett.\ B {\bf 739}, 125 (2014)
  [arXiv:1409.4396 [hep-th]].
  
  
\bibitem{Alencar:2014fga} 
  G.~Alencar, R.~R.~Landim, M.~O.~Tahim and R.~N.~Costa Filho,
  Phys.\ Lett.\ B {\bf 742}, 256 (2015)
  [arXiv:1409.5042 [hep-th]].
  
\bibitem{Jardim:2014cya} 
  I.~C.~Jardim, G.~Alencar, R.~R.~Landim and R.~N.~Costa Filho,
  Phys.\ Rev.\ D {\bf 91}, no. 4, 048501 (2015)
  [arXiv:1411.5980 [hep-th]].
  
\bibitem{Jardim:2014xla} 
  I.~C.~Jardim, G.~Alencar, R.~R.~Landim and R.~N.~C.~Filho,
  arXiv:1411.6962 [hep-th].
  
\bibitem{Alencar:2015awa} 
  G.~Alencar, R.~R.~Landim, C.~R.~Muniz and R.~N.~C.~Filho,
  arXiv:1502.02998 [hep-th].
  
\bibitem{Germani:2011cv} 
  C.~Germani,
  Phys.\ Rev.\ D {\bf 85}, 055025 (2012)
  [arXiv:1109.3718 [hep-ph]].
  
\bibitem{Bazeia:2005hu} 
  D.~Bazeia and L.~Losano,
  Phys.\ Rev.\ D {\bf 73}, 025016 (2006)
  [hep-th/0511193].

\bibitem{Bazeia:2004yw} 
  D.~Bazeia, F.~A.~Brito and A.~R.~Gomes,
  JHEP {\bf 0411}, 070 (2004)
  [hep-th/0411088].

\bibitem{Bazeia:2003qt} 
  D.~Bazeia, J.~Menezes and R.~Menezes,
  Phys.\ Rev.\ Lett.\  {\bf 91}, 241601 (2003)
  [hep-th/0305234].

\bibitem{Liu:2009ve} 
  Y.~X.~Liu, J.~Yang, Z.~H.~Zhao, C.~E.~Fu and Y.~S.~Duan,
  Phys.\ Rev.\ D {\bf 80}, 065019 (2009)
  [arXiv:0904.1785 [hep-th]].


\bibitem{Zhao:2009ja} 
  Z.~H.~Zhao, Y.~X.~Liu and H.~T.~Li,
  Class.\ Quant.\ Grav.\  {\bf 27}, 185001 (2010)
  [arXiv:0911.2572 [hep-th]].

\bibitem{Liang:2009zzf} 
  J.~Liang and Y.~S.~Duan,
  Phys.\ Lett.\ B {\bf 681}, 172 (2009).


\bibitem{Zhao:2010mk} 
  Z.~H.~Zhao, Y.~X.~Liu, H.~T.~Li and Y.~Q.~Wang,
  Phys.\ Rev.\ D {\bf 82}, 084030 (2010)
  [arXiv:1004.2181 [hep-th]].

\bibitem{Zhao:2011hg} 
  Z.~H.~Zhao, Y.~X.~Liu, Y.~Q.~Wang and H.~T.~Li,
  JHEP {\bf 1106}, 045 (2011)
  [arXiv:1102.4894 [hep-th]].






\bibitem{Ahmed:2013lea} 
  A.~Ahmed, B.~Grzadkowski and J.~Wudka,
  JHEP {\bf 1404}, 061 (2014)
  [arXiv:1312.3576 [hep-th]].
  
\bibitem{Guo:2011wr} 
  H.~Guo, Y.~X.~Liu, Z.~H.~Zhao and F.~W.~Chen,
  Phys.\ Rev.\ D {\bf 85}, 124033 (2012)
  [arXiv:1106.5216 [hep-th]].

\bibitem{Dzhunushaliev:2009va} 
  V.~Dzhunushaliev, V.~Folomeev and M.~Minamitsuji,
  Rept.\ Prog.\ Phys.\  {\bf 73}, 066901 (2010)
  [arXiv:0904.1775 [gr-qc]].
  
\bibitem{Movahed:2007ps} 
  M.~S.~Movahed and S.~Ghassemi,
  Phys.\ Rev.\ D {\bf 76}, 084037 (2007)
  [arXiv:0705.3894 [astro-ph]].

\bibitem{Du:2013bx} 
  Y.~Z.~Du, L.~Zhao, Y.~Zhong, C.~E.~Fu and H.~Guo,
  Phys.\ Rev.\ D {\bf 88}, 024009 (2013)
  [arXiv:1301.3204 [hep-th], arXiv:1301.3204 [hep-th]].
  

\bibitem{Gremm:1999pj} 
  M.~Gremm,
  Phys.\ Lett.\ B {\bf 478}, 434 (2000)
  [hep-th/9912060].

\bibitem{Bazeia:2004dh} 
  D.~Bazeia and A.~R.~Gomes,
  JHEP {\bf 0405}, 012 (2004)
  [hep-th/0403141].

\bibitem{Bazeia:2006ef} 
  D.~Bazeia, F.~A.~Brito and L.~Losano,
  JHEP {\bf 0611}, 064 (2006)
  [hep-th/0610233].
\bibitem{Csaki:2000fc} 
  C.~Csaki, J.~Erlich, T.~J.~Hollowood and Y.~Shirman,
  Nucl.\ Phys.\ B {\bf 581}, 309 (2000)
  [hep-th/0001033].

\bibitem{German:2012rv} 
  G.~German, A.~Herrera-Aguilar, D.~Malagon-Morejon, R.~R.~Mora-Luna and R.~da Rocha,
  JCAP {\bf 1302}, 035 (2013)
  [arXiv:1210.0721 [hep-th]].
  
\bibitem{Landim:2011ki} 
  R.~R.~Landim, G.~Alencar, M.~O.~Tahim and R.~N.~Costa Filho,
  JHEP {\bf 1108}, 071 (2011)
  [arXiv:1105.5573 [hep-th]].
  
\bibitem{Landim:2011ts} 
  R.~R.~Landim, G.~Alencar, M.~O.~Tahim and R.~N.~Costa Filho,
  JHEP {\bf 1202}, 073 (2012)
  [arXiv:1110.5855 [hep-th]].
 
 
\bibitem{Alencar:2012en} 
  G.~Alencar, R.~R.~Landim, M.~O.~Tahim and R.~N.~C.~Filho,
  JHEP {\bf 1301}, 050 (2013)
  [arXiv:1207.3054 [hep-th]].
  

\bibitem{Cvetic:2008gu} 
  M.~Cvetic and M.~Robnik,
  Phys.\ Rev.\ D {\bf 77}, 124003 (2008)
  [arXiv:0801.0801 [hep-th]].

\bibitem{Landim:2013dja} 
  R.~R.~Landim, G.~Alencar, M.~O.~Tahim and R.~N.~Costa Filho,
  Phys.\ Lett.\ B {\bf 731}, 131 (2014)
  [arXiv:1310.2147 [hep-th]].

\bibitem{Melfo:2002wd} 
  A.~Melfo, N.~Pantoja and A.~Skirzewski,
  Phys.\ Rev.\ D {\bf 67}, 105003 (2003)
  [gr-qc/0211081].

 


  \end{thebibliography}
\end{document}